\documentclass[aps,prev,twocolumn,preprintnumbers,floatfix,nofootinbib]{revtex4-1}

\usepackage{graphicx}
\usepackage[ngerman,english]{babel}
\usepackage{bm}
\usepackage{times}
\usepackage{slashed}
\usepackage{mathtools}
\usepackage{color}
\usepackage{epsfig}
\usepackage{dcolumn}
\usepackage{textcomp}
\usepackage{color}
\newcommand\scalemath[2]{\scalebox{#1}{\mbox{\ensuremath{\displaystyle #2}}}}

\begin{document} 

\preprint{IIPDM-2019}

\title{A Two Higgs Doublet Model for Dark Matter and Neutrino Masses}

\author{Daniel A. Camargo$^1$}\email{daniel.camargo@iip.ufrn.br}
\author{Miguel D. Campos$^{2,3}$}
\author{Tessio B. de Melo$^{1,4}$}
\author{Farinaldo S. Queiroz$^{1}$}

\affiliation{$^1$International Institute of Physics, Universidade Federal do Rio Grande do Norte, Campus Universitario, Lagoa Nova, Natal-RN 59078-970, Brazil}
\affiliation{$^2$ Department of Physics, King’s College London, Strand, London WC2R 2LS, UK}
\affiliation{$^3$ Max-Planck-Institut für Kernphysik, Saupfercheckweg 1, 69117 Heidelberg, Germany}
\affiliation{$^4$Departamento de F\'isica, Universidade Federal da Para\'iba, Caixa Postal 5008, 58051-970, Jo\~ao Pessoa-PB, Brazil.}

\begin{abstract}
\noindent
Motivated by the interesting features of Two Higgs Doublet Models (2HDM) we present a 2HDM extension where the stability of dark matter, neutrino masses and the absence of flavor changing interactions are explained by promoting baryon and lepton number to gauge symmetries. Neutrino masses are addressed within the usual type I seesaw mechanism. A vector-like fermion acts as dark matter and it interacts with Standard Model particles via the kinetic and mass mixings between the neutral gauge bosons. We compute the relevant observables such as the dark matter relic density and spin-independent scattering cross section to outline the region of parameter space that obeys current and projected limits from collider and direct detection experiments via thermal and non-thermal dark matter production. 
\end{abstract}

\maketitle
\flushbottom

\section{Introduction}
\label{sec_introduction}

The Standard Model (SM) constitutes the most accurate description of the electroweak and strong interactions in nature \cite{Glashow:1961tr,Weinberg:1967tq,Tanabashi:2018oca}. Its success is remarkable and has endured a wealth of experimental scrutiny over the past decades. The observation of the so called Higgs boson \cite{Aad:2012tfae,Chatrchyan:2012xdj} with interactions that resemble those predicted by the SM further supports its consistency. We have observed no robust signs of physics beyond the SM at the LHC thus far. Moreover, the $\rho$ parameter, $\rho=m_W/(m_Z \cos \theta_W)$, which is equal to one in the SM \cite{Tanabashi:2018oca}, has been precisely measured to be indeed close to unit. These facts have constrained several extended scalar sectors. That being said, it is natural to conceive the existence of additional scalars since scalar doublets with weak hyperchage equal to $\pm 1$ and neutral scalar singlets do not perturb the $\rho$ parameter and scalar masses above 500 GeV are just now starting to be probed at the LHC.\\

Moreover, we have more fundamental reasons to foresee physics beyond the SM. The SM does not explain neutrino masses and the presence of dark matter in our universe \cite{Bertone:2016nfn}.  Non-zero neutrino masses have been conclusively established via the observation of neutrino oscillations \cite{Fukuda:1998mi,Apollonio:1999ae} and the presence of a non-baryonic dark matter component in our universe has been confirmed through a variety of cosmological observations \cite{Bertone:2016nfn}. \\

Initially 2HDM surfaced because they naturally keep the $\rho$ parameter unaltered \cite{Lee:1973iz}, resemble the scalar sector of the Supersymmetric Standard Model \cite{Gunion:2002zf}, and give rise to interesting  collider \cite{Davidson:2010sf,Nomura:2017wxf,Camargo:2018klg} and astrophysical studies \cite{Turok:1990zg,Cline:1995dg,Clarke:2015hta}. These models simply add to the SM spectrum an extra scalar doublet that may contribute to the fermion masses. The possible forms of generating fermion masses give rise to different classes of 2HDM. Despite these nice features, 2HDM do not address the two aforementioned phenomena and are plagued with flavor changing interactions. Thus, if they are meant to represent a road towards beyond the SM, they should somehow accommodate neutrino masses and dark matter. Several extensions of the 2HDM have been proposed trying to accommodate neutrino masses \cite{Antusch:2001vn,Atwood:2005bf,Liu:2016mpf,Cheung:2017lpv,Arcadi:2017wqi,Bertuzzo:2018ftf}, dark matter \cite{LopezHonorez:2006gr,Gustafsson:2007pc,Dolle:2009fn,Chao:2012pt,Goudelis:2013uca,Honorez:2010re,LopezHonorez:2010tb,Arhrib:2013ela,Bonilla:2014xba,Queiroz:2015utg,Arcadi:2018pfo}, and the absence of flavor changing interactions \cite{Ma:1998dx,Ma:2000cc,Ma:2002nn,Grimus:2009mm}.  It would be interesting if these issues were to be addressed via gauge principles in this context, motivating further models
 \cite{Huang:2015wts,Arhrib:2018sbz,Heeck:2014qea,Crivellin:2015mga,DelleRose:2017xil}. In particular, some could simultaneously explain the absence of flavor changing interactions  and neutrino masses \cite{Ko:2012hd,Ko:2013zsa,Ko:2014uka,Ko:2015fxa,Campos:2017dgc}.  \\
 
In this work, we give a step further and discuss an extension of the 2HDM where neutrino masses, the absence of flavor changing interactions and dark matter are simultaneously addressed by promoting the baryon and lepton number to gauge symmetries and adding a vector-like fermion as dark matter. The two scalar doublets have different charges under the $U(1)_{B-L}$ and for this reason only one will be able to generate the SM fermion masses, which will automatically avoid flavor changing interactions. Furthermore, the presence of this new abelian gauge symmetry requires the addition of three right-handed neutrinos to cancel the gauge anomalies, and they will mix with the active neutrinos to generate the active neutrino masses via the type I seesaw mechanism \cite{Minkowski:1977sc,Mohapatra:1979ia,Schechter:1980gr}. In this way, one gauge symmetry is responsible for stabilizing the dark matter particle, prohibiting flavor changing interactions, and explaining neutrino masses via the seesaw mechanism.\\

In order to assess whether our assumptions generate a viable model we compute the dark matter relic density within the usual thermal freeze-out and the dark matter-nucleon scattering cross section, to later face them against the existing and projected limits from direct detection experiments. For completeness we also include the subdominant bounds stemming from LHC searches.  We also pursue the scenario of late entropy injection as a mechanism to ameliorate the direct detection bounds. \\

Our work is structured as follows: In {\it Section} \ref{sec_model} we review the 2HDM-$U(1)_{B-L}$ model; in {\it Section} \ref{sec_collider} we discuss the dark matter observables and collider limits. In {\it Section} \ref{sec_conclusion} we draw our conclusions.

\section{The Model}
\label{sec_model}
\begin{table*}[!t]
\centering
\begin{tabular}{ccccccccccc}
\hline 
Fields & $u_R$ & $d_R$ & $Q_L$ & $L_L$ & $e_R$ & $N_R$ & $\Phi _2$  & $\Phi_1$ & $\Phi_S$ & $\chi_{L,R}$ \\ \hline 
Charges & $u$ & $d$ & $\frac{(u+d)}{2}$ & $\frac{-3(u+d)}{2}$ & $-(2u+d)$ & $-(u+2d)$ & $\frac{(u-d)}{2}$ & $\frac{5u}{2} +\frac{7d}{2}$ & $2u+4d$ & $Q_\chi$ \\
$U(1)_{B-L}$ & $1/3$ & $1/3$ & $1/3$ & $-1$ & $-1$ & $-1$ & $0$ & $2$ & 2 & $Q_\chi$\\
\hline
\end{tabular}
\caption{$U(1)_{B-L}$-charges for all the fermions and scalars of the model. In particular, this assignment of charges is able to explain neutrino masses and the absence of flavor changing interactions in the type I 2HDM.}
\label{cargas_u1_2hdm_tipoI}
\end{table*}
Motivated by the lack of neutrino masses and dark matter in the usual 2HDM and the interesting phenomelogical consequences of gauging lepton and baryon numbers, we present a 2HDM-$U(1)_{B-L}$ model, where neutrino masses, dark matter, and the absence of flavor changing interactions are simultaneously addressed. The full model content is summarized in {\it Table}\,\ref{cargas_u1_2hdm_tipoI}. Setting neutrino masses aside, our model is a type I 2HDM, where only one scalar doublet contributes to SM fermion masses via the Yukawa lagrangian below,
\begin{equation}
\begin{split}
\label{2hdm_tipoI_u1}
\mathcal{L} _{Y _{\text{2HDM}}} &= y_2 ^d \bar{Q} _L \Phi _2 d_R + y_2 ^u \bar{Q} _L \widetilde \Phi _2 u_R + y_2 ^e \bar{L} _L \Phi _2 e_R \\
&+ y^{D} \bar{L} _L \widetilde \Phi _2 N_R + Y^{M} \overline{(N_R)^{c}}\Phi_{s}N_R + h.c. \\
\end{split}
\end{equation}where the scalar doublets are written as, 
\begin{equation}
\Phi _i = \begin{pmatrix} \phi ^+ _i \\ \left( v_i + \rho _i + i\eta _i \right)/ \sqrt{2}\end{pmatrix},
\end{equation}and the singlet scalar as $\Phi_s = (\phi_s + v_s + i I_\phi)/\sqrt{2}$.
\\

Through Eq.\ \eqref{2hdm_tipoI_u1} we can generate masses to all fermions after the spontaneous symmetry breaking mechanism takes place. The last term in Eq.\ \eqref{2hdm_tipoI_u1} is a new addition to the usual type I 2HDM, which arises due to the presence of right-handed neutrinos as required by the $U(1)_{B-L}$ gauge symmetry \footnote{Obviously there are ways to cancel the gauge anomalies without three right-handed neutrinos as explored in \cite{Patra:2016ofq,Bernal:2018aon}.}. With these right-handed neutrinos the type I seesaw mechanism is realized, yielding active and right-handed neutrino masses given respectively by $ m _\nu = - m _D ^T M _R ^{-1} M _D $ and $ m _N = M _R $, where $ m _D = y ^D v _2 / 2 \sqrt{2} $ and $ M _R = y ^M v _s / 2 \sqrt{2} $, with $M _R \gg m _D$. The scalar singlet $\Phi_s$ is responsible for breaking $B-L$ at sufficiently high scales. The charges of the fields under $U(1)_{B-L}$ are shown in {\it Table}\, \ref{cargas_u1_2hdm_tipoI}.  The fact that $\Phi_1$ and $\Phi_2$ transform differently under $B-L$ prohibits one of them from generating fermion masses, consequently avoiding flavor changing neutral interactions. \\

The scalars that generate fermion masses give rise to the following scalar potential,  \\

\begin{equation}
\begin{split}
V &= m_{11} ^2 \Phi _1 ^\dagger \Phi _1 + m_{22} ^2 \Phi _2 ^\dagger \Phi _2 + \frac{\lambda _1}{2} \left( \Phi _1 ^\dagger \Phi _1 \right) ^2 + \frac{\lambda _2}{2} \left( \Phi _2 ^\dagger \Phi _2 \right) ^2  \\
&+ \lambda _3 \left( \Phi _1 ^\dagger \Phi _1 \right) \left( \Phi _2 ^\dagger \Phi _2 \right) + \lambda _4 \left( \Phi _1 ^\dagger \Phi _2 \right) \left( \Phi _2 ^\dagger \Phi _1 \right) \\
& + m_s ^2 \Phi _s ^\dagger \Phi _s + \frac{\lambda _s}{2} \left( \Phi _s ^\dagger \Phi _s \right) ^2 + \lambda _{s1} \Phi _1 ^\dagger \Phi _1 \Phi _s ^\dagger \Phi _s \\
&+ \lambda _{s2} \Phi _2 ^\dagger \Phi _2 \Phi _s ^\dagger \Phi _s - \left( \mu \Phi _1 ^\dagger \Phi _2 \Phi _s + h.c. \right).\\
\end{split}
\label{pot_2hdm_U1}
\end{equation}

Similarly to fermions, after spontaneous symmetry breaking, gauge bosons masses are also generated. The gauge symmetries are spontaneously broken after $\Phi_1, \Phi_2$ and $\Phi_s$ acquire vacuum expectations values $v_1, v_2$ and $v_s$ respectively. These scalar multiplets yield five physical scalar fields: 3 CP-even, $H$, $H _s$ and $h$. The latter being the Higgs with $m_h=125$ GeV. A CP-odd scalar $A$ and charged scalar $H ^+$ are also present in the spectrum similarly to the usual 2HDM.\\

The fully analytic expressions for the scalar masses are lengthy and have been derived in \cite{Campos:2017dgc}. As for the masses of $A$ and $H ^+$ scalar they read, 
\begin{equation}
m _A ^2 = \frac{\mu ( v _1 ^2 v _2 ^2 + v ^2 v _s ^2 )}{\sqrt{2} v _1 v _2 v _s} ,
\label{pseudoscalar_mass}
\end{equation}
\begin{equation}
m _{H ^+} ^2 = \frac{( \sqrt{2} \mu v _s - \lambda _4 v _1 v _2 ) v ^2}{2 v _1 v _2} ,
\label{charged_scalar_mass}
\end{equation}
with $v ^2 = v _1 ^2 + v _2 ^2 = 246 ^2 \text{GeV} ^2$. The positiveness of $m _A ^2$ restricts $\mu$ and $v _s$ to have the same sign, which we choose to be positive. Similarly, from $m _{H ^+} ^2$ we see that the parameter $\mu$ is bounded from below, 
\begin{equation}
\mu > \mu _{\text{min}} = \frac{\lambda _4 v _1 v _2}{\sqrt{2} v _s} .
\end{equation}

We assume here that $v _s \gg v$, which requires small $\mu$ values. For instance, taking $\lambda _4 = 0.1$, $v _1 = v _2 = 174$ GeV, and $v _s = 10$ TeV, we obtain $\mu _{\text{min}} = 215$ MeV. We highlight that these choices for the couplings are fully consistent with the stability of the scalar potential \cite{Xu:2017vpq,Chen:2018uim}.\\

The model consists of a $B-L$ gauge extension containing two $SU(2)$-scalar doublets and a singlet. In addition, a vector-like fermion plays the role of dark matter. The dark matter phenomenology is governed by the $Z^\prime$ boson that stems from the $U(1)_{B-L}$ symmetry. To avoid new anomalies, the dark matter candidate cannot be chiral under $B-L$ and its charges are such that $Y_{\chi_L}=Y_{\chi_R}$. This defines a dark fermion Lagrangian with a bare-mass term $m_\chi\bar \chi \chi$ leading to,
\begin{align}
\mathcal{L}_\text{DM}= i \bar \chi \slashed D \chi - m_\chi \bar \chi \chi \; ,
\end{align}
with the covariant derivative of the SM-singlet fermion $D_\mu = \partial_\mu -ig_{\chi} \hat Z'_\mu$. We will set $g_{\chi}=g_{BL}Y_{\chi}$ throughout this analysis, and use $g_{\chi}$ as a free parameter. Here $\hat{Z^\prime}$ is in the non-physical basis. In our model, the $\hat Z$ and $\hat Z^\prime$ gauge bosons will mix with one another. After the diagonalization procedure, we will find the SM $Z$ and a massive $Z^\prime$ boson as mass eigenstates. As a result of this mixing, the dark matter fermion will interact with the $Z$ boson, and consequently to all SM fermions. In summary, the dark matter fermion will interact with all fermions charged under $B-L$ via the $Z^\prime$ boson, and to all SM fermions via a $Z$ exchange. Moreover, the scalar doublets are charged under $B-L$, so they will introduce additional interactions involving the $Z^\prime$ boson. For these reasons our results can differ from those obtained in studies involving simplified models \cite{Alves:2013tqa,Arcadi:2013qia,Alves:2015pea,Camargo:2018rfi,Arcadi:2017kky}. 

Our study is based on gauge invariance and for this reason we have also included a kinetic mixing term, $\epsilon$, between the field strength tensors of $U(1)_{B-L}$ and $U(1)_Y$,
\begin{equation}
\mathcal{L} _{\rm gauge} =  - \frac{1}{4} B _{\mu \nu} B^{\mu \nu} + \frac{\epsilon}{2\, \cos \theta_W} X _{\mu \nu} B^{\mu \nu} - \frac{1}{4} X _{\mu \nu} X ^{\mu \nu} ,
\label{Lgaugemix1}
\end{equation}
where $X _{\mu \nu}$ is the field strength tensor of the new symmetry. This extra kinetic term will generate a mixing between the neutral gauge bosons. 
We highlight that the neutral gauge bosons will mix due to the kinetic and the mass mixing terms. The latter arises via the scalar doublet $\Phi_1$, which is charged under $B-L$, and contributes to the $Z^\prime$ mass (See Appendix for an explicit proof). \\ 


In order to bring the kinetic Lagrangian \eqref{Lgaugemix1} to the canonical form a $GL(2,R)$-rotation is required, which is followed by the electroweak rotation, making the photon $A _\mu$ decouple from the massive gauge bosons $Z ^0 _\mu$ and $X _\mu$, which are still mixed. We can then obtain the physical eigenstates as a linear combination of these gauge eigenstates further rotating them by an angle $\xi$ as follows

\begin{equation}
\label{rotacao_zz_fisicos}
\begin{pmatrix} Z_\mu \\ Z ' _\mu \end{pmatrix} = \begin{pmatrix} \cos \xi & - \sin \xi \\ \sin \xi & \cos \xi \end{pmatrix} \begin{pmatrix} Z^0 _\mu \\ X_\mu \end{pmatrix}
\end{equation}
where the dependence of the mixing angle with the parameters of the model is given in the Appendix (see Eq.\ \eqref{Eq:xi}). After having described the model and the particle spectrum we will discuss the dark matter and collider observables.

\section{Collider Bounds}

\begin{figure}[h!]
\includegraphics[width=1\columnwidth]{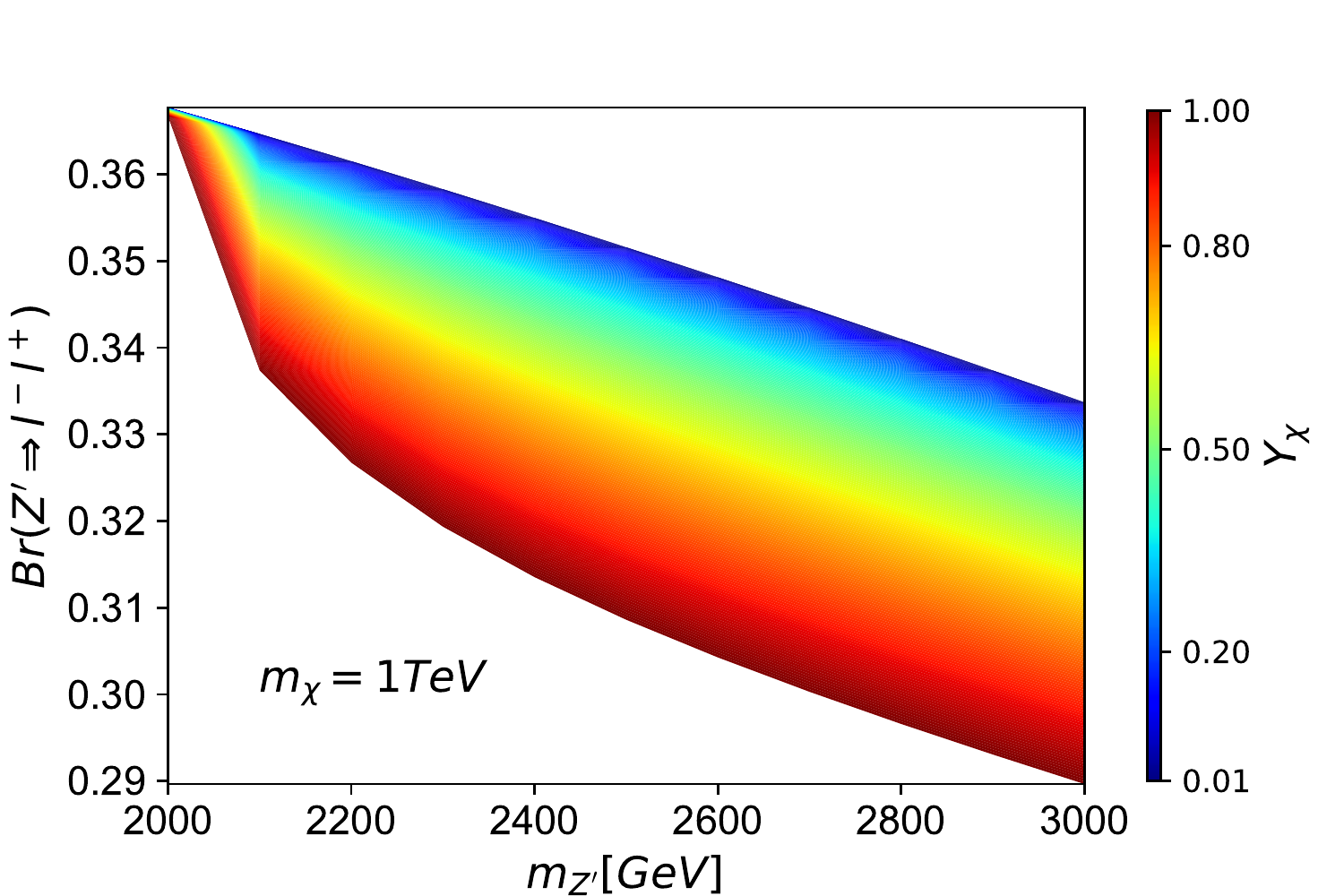}
\caption{Dilepton branching ratio ($l = e, \mu$) as function of $m_{Z^\prime}$ and $Y_\chi$ for $m_\chi = 1$ TeV.}
\label{fig:BR}
\end{figure}

The most relevant collider bounds applicable to our model stem from LHC searches for new heavy resonances. These searches rely on the narrow width approximation. It has been shown elsewhere that for $g_{BL}>0.4$ the narrow width approximation is violated \cite{Camargo:2018klg}. Therefore to be conservative we adopted $g_{BL}=0.1$ throughout. These searches look for excess events in the dilepton channel ($ee,\mu\mu$) that could be produced via the resonant production of a $Z^\prime$ gauge boson such as ours. Without dark matter, these limits constrain at the end of the day the $g_{BL}$ coupling and the $Z^{\prime}$ mass which control the $Z^\prime$ production cross section at the LHC and its branching ratio into dileptons. With the presence of dark matter the lower mass limit can be weakened if the decay into invisible (dark matter) is open. That happens when $m_{Z^\prime} > 2 m_{\chi}$. The larger $g_{\chi}$ the larger the branching ratio into dark matter. In other words, as we increase $g_{\chi}$ the LHC limit on the $Z^\prime$ mass weakens. However, the impact of the dark channel on the bounds depends at the end on how much the dilepton branching ratio changes when the dark channel is opened. In Fig.\ \eqref{fig:BR} we show how the branching ratio in the dilepton channel changes as the dark charge $Y_\chi$ increase assuming a benchmark point with $m_\chi =1$ TeV and the scalar masses in the TeV scale as is depicted by the table \eqref{param}. In this case, there is a slight change and the LHC bounds do not suffer considerable changes as is clearly noticed in Fig.\ \ref{fig:relics}. Therefore the LHC bounds are subdominant in our model, but important because offer an orthogonal cross check to it. We will now discuss the dark matter observables.

\section{THERMAL PRODUCTION AND NON-THERMAL PRODUCTION}
\label{sec_collider}
The nature of dark matter is one of the greatest mysteries in science today \cite{Queiroz:2016sxf,Catena:2017xqq,Queiroz:2018utk}. We know that dark matter accounts for $27\%$ of the energy budget of our universe. That translates into $\Omega h^2 =0.11$, where h is the error encoded in the Hubble rate measurement. It is desirable that any UV complete model that attempts to replace the SM should at least explain neutrino masses and dark matter. We already discussed how neutrino masses are accommodated in our model, and now we will address the presence of dark matter in our universe through a dark fermion. In this work, we will conceive two different dark matter production mechanism:(i) thermal production via freeze-out; (ii) non-thermal production via late-time entropy injection.\\

In the standard freeze-out the dark matter abundance is found by solving the Boltzmann equation with the help of the {\it Micromegas} package \cite{Belanger:2006is,Belanger:2008sj}, after implementing the model in Feynrules \cite{Alloul:2013bka}. The dark matter abundance is set by the dark matter annihilation cross section at freeze-out. In our model, the dominant cross section is s-wave, therefore the annihilation cross section at freeze-out is equal to the annihilation cross section today.  This is important, because in this case indirect detection searches constitute powerful probes \cite{Acharya:2017ttl}. In light of the stringent bounds on light dark matter \cite{Bertuzzo:2017lwt,DeAngelis:2017gra}, we will focus on dark matter masses at the TeV scale. \\

\begin{figure*}[t!]
\includegraphics[width=0.35\columnwidth]{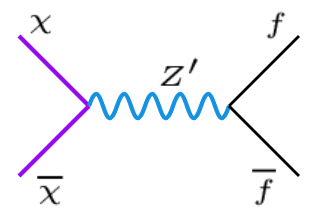}
\includegraphics[width=0.35\columnwidth]{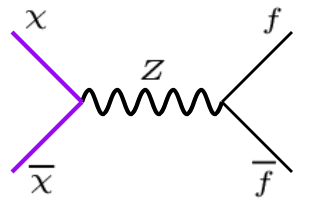}
\includegraphics[width=0.35\columnwidth]{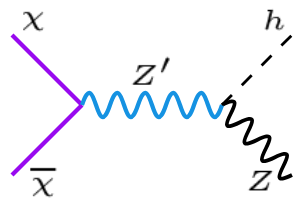}
\includegraphics[width=0.4\columnwidth]{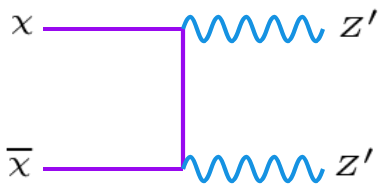}
\includegraphics[width=0.35\columnwidth]{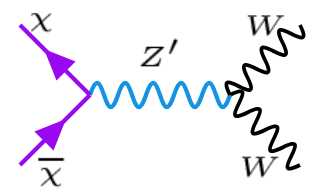}
\caption{Dark matter annihilation processes that contribute to the overall abundance. The dark matter-nucleon scattering is dictated by the t-channel version of the first diagram. There are additional channels involving the heavy higgs ($H$) and the pseudo-scalar ($A$) not shown here, but included in our numerical analysis.}
\label{fig:DMn}
\end{figure*}

In our model, several processes contribute to the overall abundance of dark matter, such as the dark matter annihilation into SM fermions via $Z^\prime$ and $Z$ s-channel exchange, the annihilation into $Z/Z^\prime,W^{\prime}$ pairs, and the one involving a $Z/Z^\prime$ in one leg and a scalar field in the other. These diagrams are displayed in Fig.\ \eqref{fig:DMn}. There are further channels involving the heavy higgs ($H$) and the pseudoscalar ($A$) not shown in Fig.\ \eqref{fig:DMn} but were included in our numerical study to precisely compute the dark matter relic density. \\

We emphasize that we will consider a vector-like fermion $\chi$ as a cold dark matter candidate whose production either simply follows a standard cosmological history or is assisted by a late time-entropy injection. In the former case, the dark matter abundance is dictated by the dark matter annihilation cross section which depends on several parameters such as $g_{\chi}$, $g_{BL}$, $\epsilon$, $m_{\chi}$ and $m_{Z^\prime}$. To reduce our free parameters, we fixed $\epsilon =10^{-3}$ in agreement with existing data \cite{Camargo:2018klg}, and set $g_{BL}=0.1$. In this way, our entire phenomenology is governed by three parameters only, $g_{\chi}$, $m_{\chi}$ and $m_{Z^\prime}$. \\

In the latter case, the dark matter at freeze-out is also set by the annihilation cross section as explained above, but its final abundance is modified due to an entropy injection episode. We will remain agnostic about the origin of this entropy injection, which can have several sources, such as late time-inflation, decays of long lived particles, and even modified expansion rate\cite{Davoudiasl:2015vba,Berlin:2016vnh,Berlin:2016gtr,DEramo:2018khz}. This late-time entropy injection will be parametrized by $\Delta$, a dilution factor. In other words, $\Omega_{DM} = \Omega_{\rm freeze-out}/\Delta$. Such episode can significantly decrease the dark matter abundance and bring an overabundant dark matter scenario back to the correct relic density. In particular, we will adopt two dilution factors, $\Delta=1,10$. Larger dilution factors are also conceivable \cite{Profumo:2003hq}. \\

That said, we computed the dark matter annihilation cross section, dark matter-nucleon spin-independent scattering cross section over a wide range of the three free parameters in our model namely, $g_{\chi}$, $m_{\chi}$, and $m_{Z^\prime}$. We emphasize that we fixed $g_{BL}=0.1$ throughout this study. \\

It is important to remember that we are dealing with a UV complete model, thus varying the $Z^\prime$ mass, also means varying the masses of the scalar particles. They are all somehow related to $v_s$ which is the scale at which the $B-L$ symmetry is spontaneously broken. In order to have a better control over the scan and grasp the physics going on in our calculations we fixed the parameter of the scalar potential, $\lambda_i$. The {\it Table} \ref{param} shows typical values of the scalar masses. \\

\begin{table}[t]
\begin{center}
\begin{tabular}{cccccc}
\hline 
$m _{Z'}$ & $v _s$ & $m _H$ & $m _{Hs}$ & $m _A$ & $m _{H ^+}$ \\ \hline 
2 & 20 & 1.01 & 6.32 & 1.02 & 1.01 \\
4 & 40 & 1.44 & 12.65 & 1.45 & 1.44 \\
6 & 60 & 1.78 & 18.97 & 1.77 & 1.77 \\
8 & 80 & 2.04 & 25.29 & 2.04 & 2.04 \\
\hline
\end{tabular}~\\
\end{center}
\caption{Physical scalar masses for $Z'$ mass values correspondent to the resonances in Figure \ref{fig:relics}. All the masses are in TeV. The parameter values used were $\lambda _1 = 0.1$, $\lambda _2 = \lambda _3 = 0.2$, $\lambda _4 = 0.38$, $\lambda _s = \lambda _{s1} = \lambda _{s2} = 0.1$, $\mu = 35$ GeV, $v _2 = 200$ GeV and $v _1 = \sqrt{v ^2 - v _2 ^2}$ with $v = 246$ GeV.}
\label{param}
\end{table}

\begin{figure*}[t!]
\includegraphics[width=1\columnwidth]{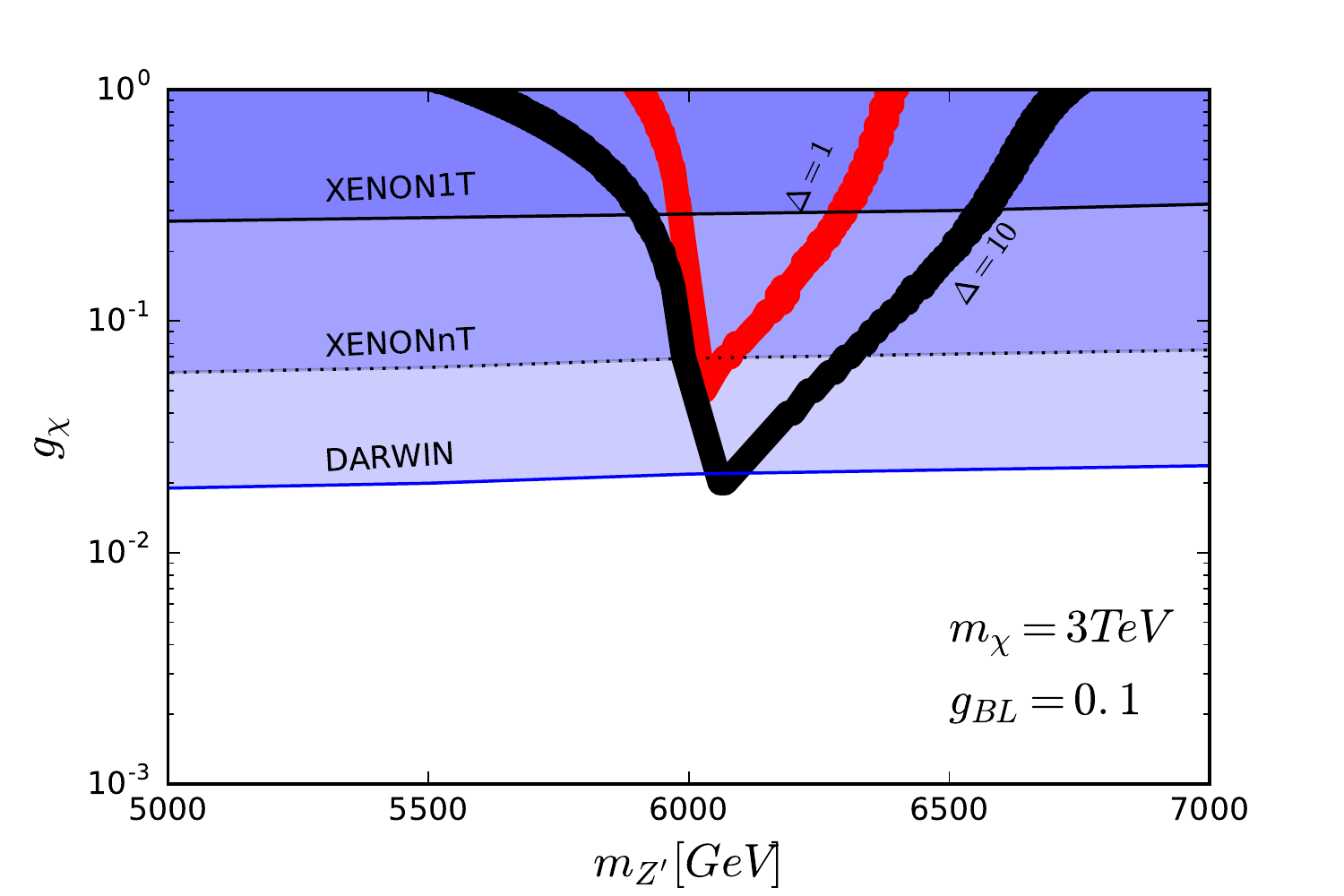}
\includegraphics[width=1\columnwidth]{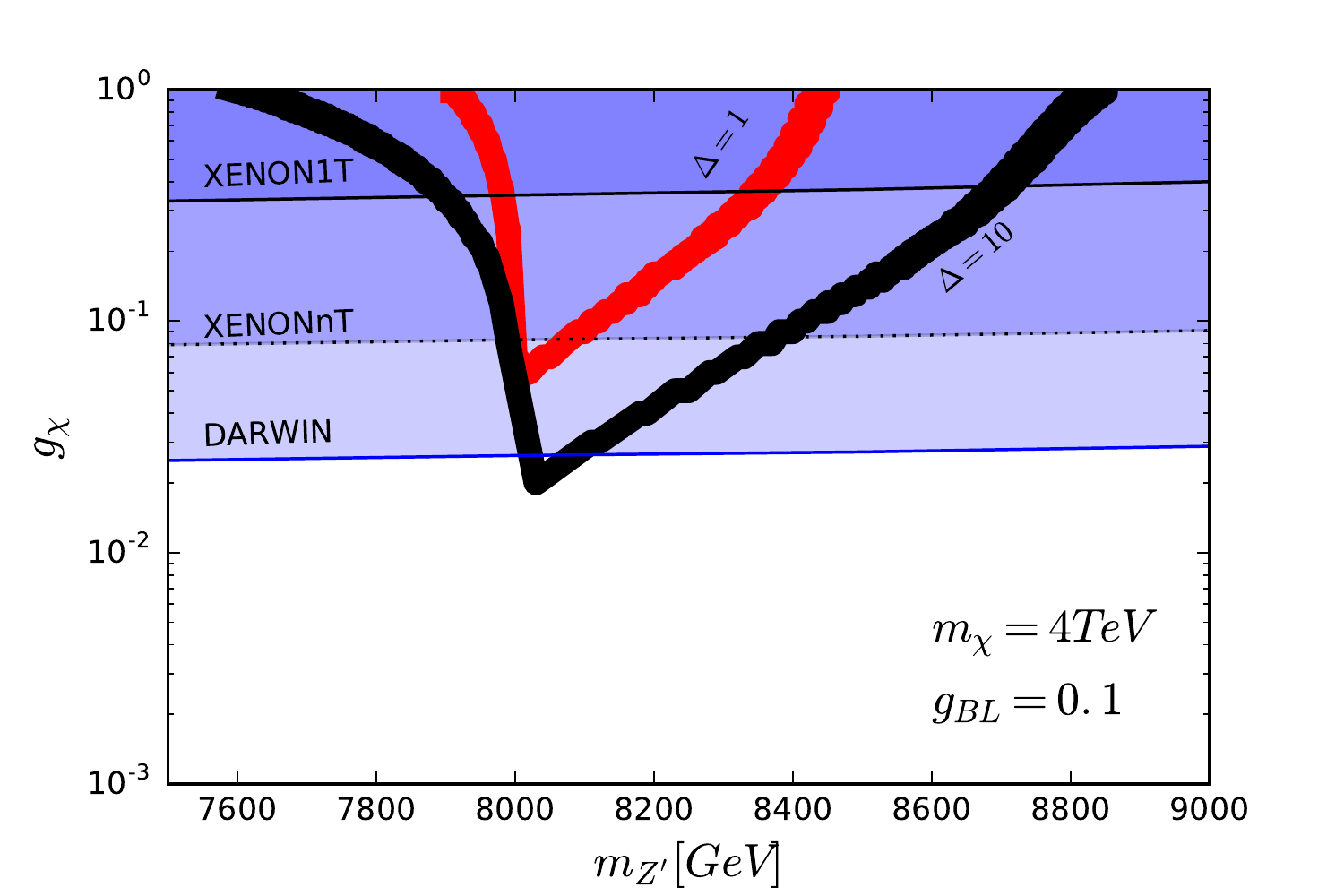}
\caption{Parameter space of the model in agreement with the measured relic abundance of the Universe for a scenario following a standard cosmology evolution $\Delta=1$ (red) and late-time inflation period with $\Delta=10$ (black). The shadow regions represents the sensitivity of the current and future experiments to the model.}
\label{fig:relics}
\end{figure*}

Concerning the relic density curves in Figure \ref{fig:relics} we show in red the parameter space that yields the correct relic density, $\Omega_\chi h^2 = 0.11$, with $\Delta=1$, i.e. in the standard cosmology case.  Notice that all panels are in the $g_{\chi}$ versus $m_{Z^\prime}$ plane. Similarly, the black line delimits the region of parameter space that reproduces the correct dark matter abundance for $\Delta=10$. As already mentioned, scenarios with larger values for $\Delta$ are conceivable, but we concentrated on these two values because they suffice for our purposes. \\

The shaded regions represent the current and projected sensitivity of few direct detection experiments. Looking at the plots one can easily conclude that even evoking non-thermal dark matter production DARWIN will thoroughly rule out dark matter masses below $1$~TeV in our model, highlighting its importance. Only if we go beyond dark matter masses of $2$~TeV, a small region of parameter space will potentially survive the projected sensitivity of DARWIN. Going for masses much above few TeV will not ameliorate the situation because the relic density curve will simply shift toward high $Z^\prime$ masses. \\

Therefore, we can conclude that future direct detection experiments will entirely probe thermal TeV scale dark matter in our model, and even scenarios that evoke for non-thermal production with a entropy injection of a factor of 10.

\begin{figure*}[t!]
\includegraphics[width=1\columnwidth]{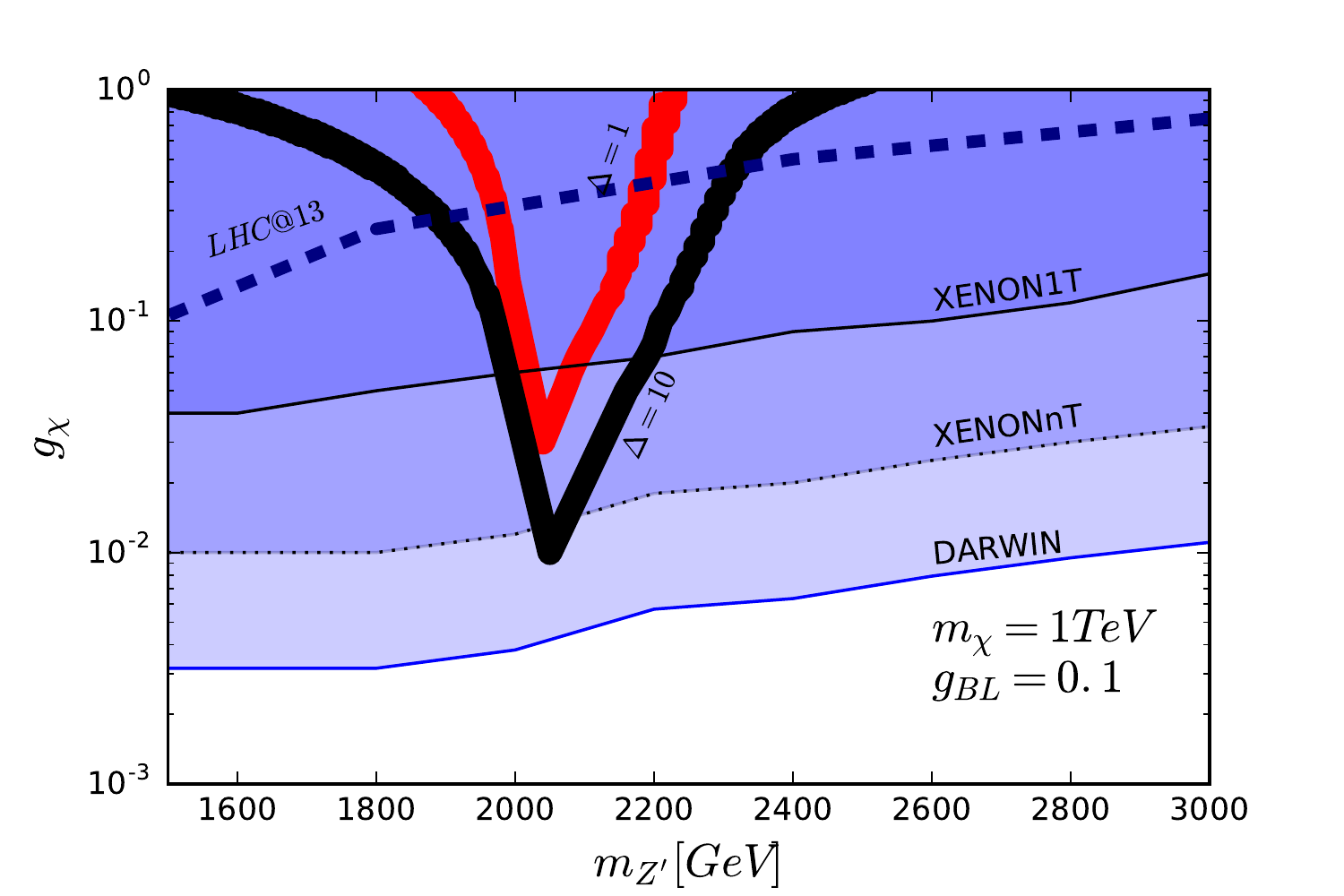}
\includegraphics[width=1\columnwidth]{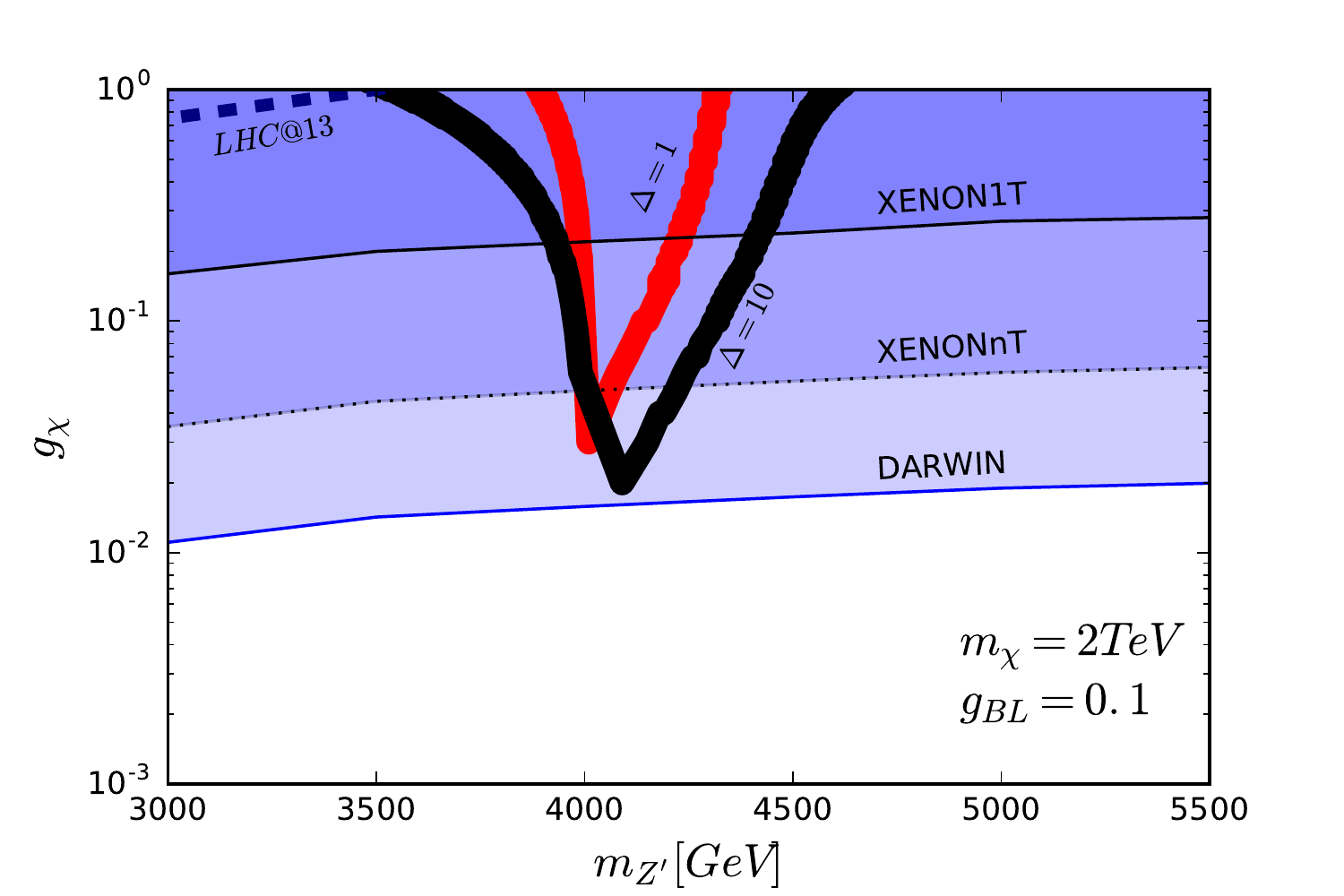}
\label{figdilepton}
\end{figure*}

\section{Conclusion}
\label{sec_conclusion}

We have discussed the phenomenology of a dark version of the canonical 2HDM. The model consists of promoting the baryon and lepton number to gauge symmetries. This symmetry is responsible for stabilizing the dark matter candidate, generating a successful seesaw mechanism and preventing flavor changing neutral interactions in our model. The anomaly cancellation requirements demands the addition of three right-handed neutrinos which are key for generating neutrino masses via a type I seesaw mechanism. A dark fermion behaves as dark matter with its phenomenology governed by few free parameters.  We concluded after considering two possible dark matter production mechanisms (thermal and non-thermal) that future direct detection experiments will basically entirely probe our model. Hence, a $B-L$ extension of the canonical 2HDM stands as a viable and interesting avenue beyond the SM since it can simultaneously address dark matter, neutrino masses and the absence of flavor changing interactions.  

\acknowledgments

The authors are thankful to Kai Schmitz and Moritz Platscher for correspondence. DC and FSQ acknowledges financial support from MEC and UFRN. TM acknowledges support from CAPES. FSQ also thanks ICTP-SAIFR FAPESP grant 2016/01343-7 for the financial support. MDC was supported by the IMPRS-PTFS and is supported  by the  European  Research Council under the European Union’s Horizon 2020 program  (ERC  Grant  Agreement  No  648680  DARK-HORIZONS).


\section*{Appendix A}\label{appendixA}

We summarize here some information aiding the understanding of the main part of the paper, by dividing it into sections relating to key computational aspects. The treatment in this appendix is valid for a general $U(1) _X$ symmetry setup, and can be specified to the $U(1) _{B-L}$ case by substituting the appropriate $Q_X$ charges as well as the gauge coupling $g _X$ by $g _{B-L}$. 

\section*{Gauge Boson Mass terms}
After rotating to a basis in which the gauge bosons have canonical kinetic terms, the covariant derivative in terms of small $\epsilon$ reads 
\begin{equation}
\label{der_cov_u1_diag}
\small
D_\mu = \partial _\mu + ig T^a W_\mu ^a + ig ' \frac{Q_Y}{2} B _{\mu} + \frac{i}{2} \left( g ' \frac{\epsilon Q_{Y}}{\cos \theta _W} + g_X Q_X \right) X_\mu ,
\end{equation}
or, explicitly,
\begin{equation}
\small
 D_\mu = \scalemath{0.8}{\partial _\mu + \frac{i}{2} \begin{pmatrix} g W_\mu ^3 + g ' Q_{Y} B_\mu + G_X X_\mu & g \sqrt{2} W_\mu ^+ \\ g \sqrt{2} W_\mu ^- & - g W_\mu ^3 + g ' Q_{Y} B_\mu + G_X X_\mu \end{pmatrix}} ,
\end{equation}
where we defined for simplicity 
\begin{equation}
\label{GXeq}
G_{Xi} = \dfrac{g ' \epsilon Q_{Y_i}}{\cos \theta _W} + g_X Q_{X_i}  , 
\end{equation}with $Q_{Y_i}$ being the hypercharge of the scalar doublet, which in the 2HDM is taken equal to $+1$ for both scalar doublets, and $Q_{X_i}$ is the charge of the scalar doublet $i$ under $U(1)_X$.\\

Then the part of the Lagrangian responsible for the gauge boson masses becomes 
\begin{equation}
\begin{split}
\mathcal{L} _{\text{mass}} = & \scalemath{0.8}{\left( D_\mu \Phi _1 \right) ^\dagger \left( D^\mu \Phi _1 \right) + \left( D_\mu \Phi _2 \right) ^\dagger \left( D^\mu \Phi _2 \right) + \left( D_\mu \Phi _S \right) ^\dagger \left( D^\mu \Phi _S \right)}\\
 = & \scalemath{0.8}{\frac{1}{4} g^2 v ^2 W_\mu ^- W ^{+ \mu} + \frac{1}{8} g_Z ^2 v ^2 Z_\mu ^0 Z^{0 \mu} - \frac{1}{4} g_Z \left( G_{X1} v_1 ^2 + G_{X2} v_2 ^2 \right) Z_\mu ^0 X ^\mu}  \\
&+ \frac{1}{8} \left( v_1 ^2 G_{X1} ^2 + v_2 ^2 G_{X2} ^2 + v_S ^2 g_X ^2 q_X ^2 \right) X_\mu X ^\mu,
\end{split}
\label{mixinggaugebosons}
\end{equation}
where $v ^2 = v_1 ^2 + v_2 ^2$. Eq. \eqref{mixinggaugebosons} can then be written as
\begin{equation}
\begin{split}
\mathcal{L} _{\rm mass} &= \scalemath{0.8}{m_W ^2 W_\mu ^- W ^{+ \mu} + \frac{1}{2} m_{Z^0} ^2 Z_\mu ^0 Z^{0 \mu} - \Delta ^2 Z_\mu ^0 X ^\mu + \frac{1}{2} m_X ^2 X_\mu X ^\mu},
\end{split}
\end{equation}
with
\begin{equation}
m_W ^2 = \frac{1}{4} g^2 v ^2, \qquad m_{Z} ^2 = \frac{1}{4} g_Z ^2 v ^2,
\label{Eq:MZ0}
\end{equation}
\begin{equation}
\Delta ^2 = \frac{1}{4} g_Z \left( G_{X1} v_1 ^2 + G_{X2} v_2 ^2 \right),
\label{Eq:Delta2}
\end{equation}
and
\begin{equation}
m_X ^2 = \frac{1}{4} \left( v_1 ^2 G_{X1} ^2 + v_2 ^2 G_{X2} ^2 + v_S ^2 g_X ^2 q_X ^2 \right) .
\label{Eq:MX}
\end{equation}
Summarizing, after the symmetry breaking, one can realize that there is a remaining mixing between $Z^0 _\mu$ and $X_\mu$ that can be expressed through the symmetric matrix
\begin{equation}
m_{Z^0X} ^2 = \frac{1}{2} \begin{pmatrix} m_{Z^0} ^2 & - \Delta ^2 \\ - \Delta ^2 & m_X ^2 \end{pmatrix}, 
\end{equation}
or, explicitly,
\begin{equation}
m_{Z^0X} ^2 = 
\scalemath{0.8}{\frac{1}{8} \begin{pmatrix} g_Z ^2 v ^2 & - g_Z \left( G_{X1} v_1 ^2 + G_{X2} v_2 ^2 \right) \\ - g_Z \left( G_{X1} v_1 ^2 + G_{X2} v_2 ^2 \right)  & v_1 ^2 G_{X1} ^2 + v_2 ^2 G_{X2} ^2 + v_S ^2 g_X ^2 q_X ^2 \end{pmatrix}} \,.
\label{mixinzx}
\end{equation}
The above expression, Eq.\ \eqref{mixinzx}, representing the mixing between the $Z^0 _\mu$ and $X_\mu$ bosons, is given as function of arbitrary $U(1)_{X}$ charges of doublet (or singlet) scalars. It is important to notice that, when $Q_{X1}=Q_{X2}$ and there is no singlet contribution, the determinant of the matrix Eq.\ \eqref{mixinzx} is zero.

The matrix in Eq.\ \eqref{mixinzx} is diagonalized through a rotation $O(\xi)$
\begin{equation}
\label{rotacao_zz_fisicos}
\begin{pmatrix} Z_\mu \\ Z ' _\mu \end{pmatrix} = \begin{pmatrix} \cos \xi & - \sin \xi \\ \sin \xi & \cos \xi \end{pmatrix} \begin{pmatrix} Z^0 _\mu \\ X_\mu \end{pmatrix}
\end{equation}
and its eigenvalues are
\begin{equation}
\begin{split}
\label{autovalores_matriz_zz}
m_{Z} ^2 &= \frac{1}{2} \left[ m_{Z ^0} ^2 + m_X ^2 - \sqrt{ \left( m_{Z ^0} ^2 - m_X^2 \right) ^2 + 4 \left( \Delta ^2 \right) ^2} \right], \\
m_{Z '} ^2 &= \frac{1}{2} \left[ m_{Z ^0} ^2 + m_X ^2 + \sqrt{ \left( m_{Z ^0} ^2 - m_X^2 \right) ^2 + 4 \left( \Delta ^2 \right) ^2} \right].
\end{split}
\end{equation}
The $\xi$ angle is given by
\begin{equation}
\tan \xi = \frac{ \Delta ^2}{m^2 _{Z^0} - m^2 _{X}} .
\label{Eq:xi}
\end{equation}
Since this mixing angle it supposed to be small, as  $m_{Z^\prime}^2 \gg m_Z^2$, we can use $\tan \xi \sim \sin \xi$ with
\begin{equation}
\sin \xi \simeq \frac{ G_{X1} v_1^2 + G_{X2} v_2^2}{m^2_{Z^\prime}} .
\label{Eq:xinew}
\end{equation}
We can expand this equation further to find a more useful expression. Substituting the expressions for $G_{Xi}$ and factoring out the $m_Z$ mass, we finally get

\begin{equation}
\sin \xi \simeq \frac{m_Z^2}{ m^2_{Z^{\prime}}}\left(\frac{g_X}{g_Z}(Q_{X1}\cos^2 \beta + Q_{X2}\sin^2 \beta) +\epsilon \tan\theta_W \right).\\
\label{eqsinxi}
\end{equation}

\section*{Gauge Bosons - Scalar Couplings}

In this appendix we show the relevant $Z'$ couplings to the scalars and gauge bosons, for general scalar charges. The particular $U(1) _{B-L}$ case is obtained by making $Q _{X1} = q _X = 2$, $Q _{X2} = 0$.

Trilinear $Z'$-gauge boson couplings:
\\ \\
$Z' W ^+ \partial _\mu W ^-$:
\begin{equation*}
\begin{split}
i g \cos \theta _W \sin \xi & [ ( \partial ^\mu W ^{- \nu} - \partial ^\nu W ^{- \mu} ) W _\nu ^+ \\
& - ( \partial ^\mu W ^{+ \nu} - \partial ^\nu W ^{+ \mu} ) W _\nu ^- ] Z ' _\mu
\end{split}
\end{equation*}
\\
$W ^+ W ^- \partial _\mu Z^\prime$:
\begin{equation*}
i g \cos \theta_W \sin \xi ( W_\mu ^+ W_\nu ^- - W_\nu ^+ W_\mu ^- ) \partial^{\mu} Z^{\prime \nu}
\end{equation*}
\\ \\
Trilinear $Z'$-scalars couplings:
\\ \\
$H Z' Z'$:
\begin{equation*}
\begin{split}
- \frac{1}{4} g _X ^2 q _X ^2 v _s & \cos \alpha _1 \sin \alpha _2 \cos ^2 \xi \\
+ \frac{v}{4 \cos ^2 \theta _W} [ & \Gamma _1 ^2 ( \sin \alpha \sin \alpha _1 \sin \alpha _2 + \cos \alpha \cos \alpha _2 ) \cos \beta \\
+ & \Gamma _2 ^2 ( \sin \alpha \cos \alpha _2 - \cos \alpha \sin \alpha _1 \sin \alpha _2 ) \sin \beta ]
\end{split}
\end{equation*}
\\
$h Z' Z'$:
\begin{equation*}
\begin{split}
& - \frac{1}{4} g _X ^2 q _X ^2 v _s \sin \alpha _1 \cos ^2 \xi \\
& - \frac{v}{4 \cos ^2 \theta _W} \cos \alpha _1 ( \Gamma _1 ^2 \sin \alpha \cos \beta - \Gamma _2 ^2 \cos \alpha \sin \beta ) 
\end{split}
\end{equation*}
\\
$h _s Z' Z'$:
\begin{equation*}
\begin{split}
\frac{1}{4} \{ g _X ^2 q _X ^2 v _s & \cos \alpha _1 \cos \alpha _2 \cos ^2 \xi \\
+ \frac{v}{\cos ^2 \theta _W} [ & \Gamma _1 ^2 ( \cos \alpha \sin \alpha _2 - \sin \alpha \sin \alpha _1 \cos \alpha _2 ) \cos \beta \\
+ & \Gamma _2 ^2 ( \cos \alpha \sin \alpha _1 \cos \alpha _2 + \sin \alpha \sin \alpha _2 ) \sin \beta ] \}
\end{split}
\end{equation*}
\\
$H Z Z'$:
\begin{equation*}
\begin{split}
\frac{1}{4} g _X ^2 q _X ^2 v _s \cos & \alpha _1 \sin \alpha _2 \sin ( 2 \xi ) \\
- \frac{1}{4} \frac{v}{\cos ^2 \theta _W} [ & \Omega _1 ( \cos \alpha \cos \alpha _2 + \sin \alpha \sin \alpha _1 \sin \alpha _2 ) \cos \beta \\
+ & \Omega _2 ( \sin \alpha \cos \alpha _2 - \cos \alpha \sin \alpha _1 \sin \alpha _2 ) \sin \beta ]
\end{split}
\end{equation*}
\\
$ H Z' \partial _\mu A$:
\begin{equation*}
\begin{split}
& - \frac{1}{2 \sqrt{v ^2 \sin ^2 \beta \cos ^2 \beta + v _s ^2 }} \\
& \{ g _X q _X  v \cos \alpha _1 \sin \alpha _2 \sin \beta \cos \beta \cos \xi \\
+ v _s \sec \theta _W [ & \Gamma _1 ( \cos \alpha \cos \alpha _2 + \sin \alpha \sin \alpha _1 \sin \alpha _2 ) \sin \beta \\
+ & \Gamma _2 ( \cos \alpha \sin \alpha _1 \sin \alpha _2 - \sin \alpha \cos \alpha _2 ) \cos \beta ] \}
\end{split}
\end{equation*}
\\ \\
$ h Z Z'$:
\begin{equation*}
\begin{split}
\frac{1}{4} & [ g _X ^2 q _X ^2 v _s \sin \alpha _1 \sin ( 2 \xi ) \\
& + \frac{v}{\cos ^2 \theta _W} \cos \alpha _1 ( \Omega _1 \sin \alpha \cos \beta - \Omega _2 \cos \alpha \sin \beta ) ]
\end{split}
\end{equation*}
\\ \\
$ h Z' \partial _\mu A$:
\begin{equation*}
\begin{split}
& \frac{1}{2 \sqrt{v ^2 \sin ^2 \beta \cos ^2 \beta + v _s ^2 }} [ - g _X q _X  v \sin \alpha _1 \sin \beta \cos \beta \cos \xi \\
& + \frac{v _s}{\cos \theta _W} ( \Gamma _1 \sin \alpha \cos \alpha _1 \sin \beta + \Gamma _2 \cos \alpha \cos \alpha _1 \cos \beta ) ]
\end{split}
\end{equation*}
\\ \\
$ h _s Z Z'$:
\begin{equation*}
\begin{split}
- \frac{1}{4} g _X ^2 q _X ^2 v _s \cos \alpha _1 & \cos \alpha _2 \sin ( 2 \xi ) \\
- \frac{1}{4} \frac{v}{\cos ^2 \theta _W} [ & \Omega _1 ( \cos \alpha  \sin \alpha _2 - \sin \alpha \sin \alpha _1 \cos \alpha _2 ) \cos \beta \\
+ & \Omega _2 ( \cos \alpha \sin \alpha _1 \cos \alpha _2 + \sin \alpha \sin \alpha _2 ) \sin \beta ]
\end{split}
\end{equation*}
\\ \\
$ h _s Z' \partial _\mu A $:
\begin{equation*}
\begin{split}
& \frac{1}{2 \sqrt{v ^2 \sin ^2 \beta \cos ^2 \beta + v _s ^2 }} \{ g _X q _X  v \cos \alpha _1 \cos \alpha _2 \sin \beta \cos \beta \cos \xi \\
& + \frac{v _s}{\cos \theta _W} [ \Gamma _1 ( \sin \alpha \sin \alpha _1 \cos \alpha _2 - \cos \alpha \sin \alpha _2 ) \sin \beta \\
& + \Gamma _2 ( \cos \alpha \sin \alpha _1 \cos \alpha _2 + \sin \alpha \sin \alpha _2 ) \cos \beta ] \}
\end{split}
\end{equation*}
\\ \\
$ Z' A \partial _\mu H $:
\begin{equation*}
\begin{split}
& \frac{1}{2 \sqrt{v ^2 \sin ^2 \beta \cos ^2 \beta + v _s ^2 }} \{ g _X q _X  v \cos \alpha _1 \sin \alpha _2 \sin \beta \cos \beta \cos \xi \\
& + v _s \sec \theta _W [ \Gamma _1 ( \sin \alpha \sin \alpha _1 \sin \alpha _2 + \cos \alpha \cos \alpha _2 ) \sin \beta \\
& + \Gamma _2 ( \cos \alpha \sin \alpha _1 \sin \alpha _2 - \sin \alpha \cos \alpha _2 ) \cos \beta ] \}
\end{split}
\end{equation*}
\\ \\
$ Z' A \partial _\mu h $:
\begin{equation*}
\begin{split}
& \frac{1}{2 \sqrt{v ^2 \sin ^2 \beta \cos ^2 \beta + v _s ^2 }} [ g _X q _X  v \sin \alpha _1 \sin \beta \cos \beta \cos \xi \\
& - \frac{v _s}{\cos \theta _W} ( \Gamma _1 \sin \alpha \cos \alpha _1 \sin \beta + \Gamma _2 \cos \alpha \cos \alpha _1 \cos \beta ) ]
\end{split}
\end{equation*}
\\ \\
$ Z' A \partial _\mu h _s $:
\begin{equation*}
\begin{split}
& - \frac{1}{2 \sqrt{v ^2 \sin ^2 \beta \cos ^2 \beta + v _s ^2 }} \{ g _X q _X v \cos \alpha _1 \cos \alpha _2 \sin \beta \cos \beta \cos \xi \\
& + \frac{v _s}{\cos \theta _W} [ \Gamma _1 ( \sin \alpha \sin \alpha _1 \cos \alpha _2 - \cos \alpha \sin \alpha _2 ) \sin \beta \\
& + \Gamma _2 ( \sin \alpha \sin \alpha _2 + \cos \alpha \sin \alpha _1 \cos \alpha _2 ) \cos \beta ] \}
\end{split}
\end{equation*}
\\ \\
$ Z' H ^+ W ^- $:
\begin{equation*}
\frac{1}{4} g g _X v ( Q _{X2} - Q _{X1} ) \sin ( 2 \beta ) \cos \xi
\end{equation*}
\\ \\
$ Z' H ^+ \partial _\mu H ^- $:
\begin{equation*}
\begin{split}
\frac{i}{2 \cos \theta _W} & \{ g \cos ( 2 \theta _W ) \sin \xi + [ g _X \cos \theta _W ( Q _{X1} \sin ^2 \beta + Q _{X2} \cos ^2 \beta ) \\
& + g \epsilon \tan \theta _W ] \cos \xi \}
\end{split}
\end{equation*}
\\ \\
$ Z' H ^- W ^+ $:
\begin{equation*}
\frac{1}{4} g g _X v ( Q _{X2} - Q _{X1} ) \sin ( 2 \beta ) \cos \xi
\end{equation*}
\\ \\
$ Z' H ^- \partial _\mu H ^+ $:
\begin{equation*}
\begin{split}
- \frac{i}{2 \cos \theta _W} & \{ g \cos ( 2 \theta _W ) \sin \xi + [ g _X \cos \theta _W ( Q _{X1} \sin ^2 \beta + Q _{X2} \cos ^2 \beta ) \\
& + g \epsilon \tan \theta _W ] \cos \xi \}
\end{split}
\end{equation*}
\\ \\

Where,
\begin{equation*}
\Gamma _i = \cos \xi ( g _X Q _{Xi} \cos \theta _W + g \epsilon \tan \theta _W ) - g \sin \xi ,
\end{equation*}
and,
\begin{equation*}
\begin{split}
\Omega _i = ( 2 & g ^2 \epsilon \tan \theta _W + 2 g g _X Q _{Xi} \cos \theta _W ) \cos ( 2 \xi ) \\
+ ( & 2 g g _X Q _{Xi} \epsilon \sin \theta _W - g ^2 + g _X ^2 Q _{Xi} ^2 \cos ^2 \theta _W \\
& + g ^2 \epsilon ^2 \tan ^2 \theta _W ) \sin ( 2 \xi ) .
\end{split}
\end{equation*}
\\ \\ \\ \\ \\

\bibliography{literature}

\end{document}